\documentclass[runningheads]{llncs}
\usepackage[T1]{fontenc}
\usepackage{amsmath,amssymb,amsfonts}
\usepackage{graphicx}
\usepackage{booktabs}
\usepackage{bbding}
\usepackage{makecell}
\usepackage{subfigure}
\usepackage{hyperref}

\begin{document}

\title{CT Synthesis with Conditional Diffusion Models for Abdominal Lymph Node Segmentation}

\author{Yongrui Yu\inst{1}* \and Hanyu Chen\inst{2}* \and Zitian Zhang\inst{3}* \and Qiong Xiao\inst{2}* \and Wenhui Lei\inst{1,4} \and  Linrui Dai\inst{1,4} \and Yu Fu\inst{2} \and Hui Tan\inst{2} \and Guan Wang\inst{3}\Envelope \and Peng Gao\inst{2}\Envelope \and Xiaofan Zhang\inst{1,4}\Envelope}

\institute{Shanghai Jiao Tong University, Shanghai, China \and Department of Surgical Oncology and General Surgery, Key Laboratory of Precision Diagnosis and Treatment of Gastrointestinal Tumors, Ministry of Education, The First Hospital of China Medical University, Shenyang, China \and Department of Radiology, The First Hospital of China Medical University, Shenyang, China \and Shanghai AI Laboratory, Shanghai, China}

\authorrunning{Y. Yu et al.}
\titlerunning{LN-DDPM}

\maketitle

\let\thefootnote\relax\footnotetext{* Equal contribution. Correspondence to: Guan Wang (cmuwangguan@sina.com); Peng Gao (pgao@cmu.edu.cn); Xiaofan Zhang (xiaofan.zhang@sjtu.edu.cn).}

\begin{abstract}
Despite the significant success achieved by deep learning methods in medical image segmentation, researchers still struggle in the computer-aided diagnosis of abdominal lymph nodes due to the complex abdominal environment, small and indistinguishable lesions, and limited annotated data. To address these problems, we present a pipeline that integrates the conditional diffusion model for lymph node generation and the nnU-Net model for lymph node segmentation to improve the segmentation performance of abdominal lymph nodes through synthesizing a diversity of realistic abdominal lymph node data. We propose LN-DDPM, a conditional denoising diffusion probabilistic model (DDPM) for lymph node (LN) generation. LN-DDPM utilizes lymph node masks and anatomical structure masks as model conditions. These conditions work in two conditioning mechanisms: global structure conditioning and local detail conditioning, to distinguish between lymph nodes and their surroundings and better capture lymph node characteristics. The obtained paired abdominal lymph node images and masks are used for the downstream segmentation task. Experimental results on the abdominal lymph node datasets demonstrate that LN-DDPM outperforms other generative methods in the abdominal lymph node image synthesis and better assists the downstream abdominal lymph node segmentation task.
\end{abstract}

\section{Introduction}
In the diagnosis and treatment of abdominal tumors, the occurrence of lymph node metastasis (LNM) plays a critical role in prognosis determination and therapeutic decision-making. Therefore, the segmentation of abdominal lymph nodes (LNs) based on radiological examination stands as a key challenge in the computer-aided diagnosis. In recent years, deep learning methods have achieved significant success in medical image segmentation across various imaging modalities \cite{litjens2017survey}. However, the segmentation performance for abdominal lymph nodes remains suboptimal \cite{nogues2016automatic,li2020deep,manjunatha2023lymph,tang2019ct}. This is partly due to the small size of abdominal lymph nodes, which makes them difficult to distinguish from the complex background of the abdominal cavity. Additionally, the annotated data is particularly scarce because the annotation of abdominal lymph nodes necessitates the expertise of experienced radiologists and pathologists. These issues impede the development of effective abdominal lymph node segmentation tools.

To fully leverage limited lymph node images and annotations, DRL-LNS \cite{li2020deep} combines weakly supervised learning and deep reinforcement learning (DRL) for segmenting thoracoabdominal lymph nodes. Tang \emph{et al.} \cite{tang2019ct} utilize generative models for lymph node synthesis to assist the lymph node segmentation task.
Generative models are capable of synthesizing a large quantity of synthetic data for augmenting medical image datasets.
Previous generative methods often utilize generative adversarial networks (GANs) \cite{goodfellow2014generative}, for example, Pix2pix \cite{isola2017image} for image-to-image translation and SPADE \cite{park2019semantic} for semantic image synthesis. Recently, diffusion models \cite{sohl2015deep} have attracted considerable attention due to their high-fidelity synthesis capability. Diffusion models are composed of a forward diffusion process and a backward denoising process. Denoising diffusion probabilistic models (DDPMs) \cite{ho2020denoising} are capable of generating synthetic images of high fidelity and diversity. Moreover, the training process of DDPMs exhibits better stability in comparison with that of GANs.

For the purpose of assisting abdominal lymph node segmentation task, the generation process should be controllable to obtain paired lymph node images and masks. Consequently, it is necessary to control the lymph node image generation process using lymph node masks through conditional generative models. Unconditional DDPMs generate synthetic images without control, whereas conditional DDPMs incorporate various kinds of conditions, such as texts, semantic mask, and bounding boxes, to enhance the controllability of the image generation process. Different kinds of conditioning methods have been proposed to regulate the generation process. Channel-wise concatenation \cite{dorjsembe2023conditional,waibel2023diffusion,zhuang2023semantic} is a straightforward yet effective conditioning method for generating images according to specified conditions. Semantic diffusion models (SDMs) \cite{wang2022semantic} incorporate semantic masks through spatially-adaptive normalization \cite{park2019semantic} and are often used for semantic image synthesis \cite{shrivastava2023nasdm,stojanovski2023echo,zhao2023high,du2023arsdm}. Latent diffusion models (LDMs) \cite{rombach2022high} regulate image generation process using cross-attention. Cross-attention module \cite{pinaya2022brain} offers greater flexibility, enabling the utilization of multi-modal conditions.

In addition to lymph node masks for realistic lymph node synthesis, abdominal surroundings where lymph nodes are located are also crucial. Therefore, simulating the abdominal environment is of great significance.
Echo from noise \cite{stojanovski2023echo} synthesizes cardiac ultrasound images conditioned on cardiac semantic maps.
Zhuang \emph{et al.} \cite{zhuang2023semantic} synthesize 2D abdominal CT images controlled by anatomical semantic masks.
Therefore, to mimic the surroundings of abdominal lymph nodes, anatomical structure masks are additionally introduced to control abdominal image synthesis process.

This paper employs the conditional diffusion model, namely LN-DDPM, to synthesize abdominal lymph node images. Through LN-DDPM, we generate a large amount of synthetic abdominal lymph node data to assist the downstream lymph node segmentation task. LN-DDPM is conditioned on lymph node masks and anatomical structures. Both anatomical structures and lymph node masks are conditioned using global structure conditioning mechanism, which is channel-wise concatenation, to construct abdominal image structures. Additionally, lymph node masks also function through local detail conditioning mechanism via cross-attention to better capture lymph node characteristics. Extensive experiments conducted on abdominal lymph node datasets demonstrate the effectiveness of LN-DDPM.

In conclusion, the main contributions of this work are:
\begin{itemize}
\item We develop a pipeline that incorporates a generative model and a segmentation model to enhance the segmentation performance of abdominal lymph nodes by leveraging the generative model to synthesize a diversity of abdominal lymph node data.
\item We propose LN-DDPM, a novel generative model conditioned on semantic masks. Lymph node masks are essential for generating paired lymph node data to assist the downstream segmentation task and are conditioned through both global structure conditioning via channel-wise concatenation and local detail conditioning via cross-attention.
\item To generate lymph node images with realistic abdominal surroundings, we use anatomical structures to regulate the generative model. The anatomical structures are conditioned via global structure conditioning mechanism.
\item We conduct experiments on abdominal lymph node datasets. The experimental results demonstrate that our proposed LN-DDPM outperforms other generative methods in terms of generation and segmentation performance.
\end{itemize}

\section{Related Work}

\subsection{CT Lymph Node Segmentation}
The exploration of applying deep learning methods for lymph node segmentation in CT images has great significance for disease discovery and diagnosis. We highlight deep learning-based methods for CT lymph node segmentation task.

Nogues \emph{et al.} \cite{nogues2016automatic} combine holistically-nested neural networks and structured optimization for thoracoabdominal lymph node cluster segmentation.
DRL-LNS \cite{li2020deep} combines an unsupervised segmentation network and a segmentation network with deep reinforcement learning. It segments CT slices in an unsupervised manner, generating pseudo ground-truths for training the U-Net segmentation network. Then, a DRL model is employed to optimize lymph node bounding boxes and segmentations simultaneously.
DiSegNet \cite{xu2021disegnet} proposes an efficient loss function to address the voxel class imbalances and integrates a multi-stage multi-scale dilated spatial pyramid pooling module into its encoder-decoder architecture.
Manjunatha \emph{et al.} \cite{manjunatha2023lymph} extract the volume of interest of lymph node candidates through a modified U-Net model and then execute false positive reduction.
While these methods primarily concentrate on enhancing model architectures and refining loss functions, Tang \emph{et al.} \cite{tang2019ct} leverage synthetic data generated by Pix2pix \cite{isola2017image} to train lymph node segmentation model. However, Tang \emph{et al.} solely use lymph node masks to control the Pix2pix model, resulting in the generation of synthetic images with unrealistic abdominal environment. Our proposed LN-DDPM utilizes both lymph node masks and anatomical structure masks to regulate the conditional diffusion model, enabling the generation of realistic lymph nodes and their surroundings.

\subsection{Medical Image Synthesis via Generative Models}
Unlike the natural image domain, labeled data is scarce in the medical image domain because accurately annotating medical images requires professional medical knowledge. The utilization of medical image synthesis tools can augment sample quantity, enhance sample diversity, and balance medical image datasets, ensuring stable training of downstream tasks.

Previously, generative adversarial networks \cite{goodfellow2014generative} were frequently used for medical image synthesis.
Pix2pix \cite{isola2017image} is a widely used image-to-image translation model.
Tang \emph{et al.} \cite{tang2019ct} adopt Pix2pix to learn the structure and context of lymph nodes and generate a large quantity of synthetic lymph node data.
SPADE \cite{park2019semantic} is designed for semantic image synthesis via spatially-adaptive normalization, which performs affine transformation on the semantic layout within the normalization layers.
However, the training process of GANs is plagued by instability drawbacks such as mode collapse, vanishing gradients, and non-convergence \cite{wiatrak2019stabilizing}.

Recently, denoising diffusion probabilistic models \cite{ho2020denoising} have achieved notable synthetic sample quality in both natural image domain and medical image domain.
Medical Diffusion \cite{khader2023denoising} synthesizes 3D CT and MRI scans via unconditional LDMs.
Pinaya \emph{et al.} \cite{pinaya2022brain} generate 3D brain MRIs using LDM with covariable conditions such as age, sex, and brain structure volumes.
Zhuang \emph{et al.} \cite{zhuang2023semantic} synthesize 2D abdominal CT images via semantic masks.
Echo from noise \cite{stojanovski2023echo} adopts semantic diffusion models \cite{wang2022semantic} to synthesize cardiac ultrasound images by semantic label maps.
Du \emph{et al.} \cite{du2023boosting} use ControlNet to control dermatoscopic lesion synthesis with visual and textual conditions.
Med-DDPM \cite{dorjsembe2023conditional} generates 3D brain tumor MRIs with one-hot encoded brain and tumor masks.
ArSDM \cite{du2023arsdm} introduces colonoscopy polyp image synthesis with polyp mask conditional diffusion models.
In addition to synthesizing those relatively large targets, Zhao \emph{et al.} \cite{zhao2023high} synthesize 2D CT slices of pulmonary nodules. NASDM \cite{shrivastava2023nasdm} generates nuclei pathology images via nuclei masks conditioned semantic diffusion models.

However, these methods primarily focus on generating images of large objects, most of which generate 2D images. This paper concentrates on synthesizing 3D CT images of small lymph node targets. Distinguishing abdominal lymph nodes and blood vessels from 2D images is challenging, which requires the examination of contiguous slices of 3D images to differentiate between them. This motivates the generation of 3D CT lymph node images in this study.

\subsection{Diffusion Models for Image Synthesis}
Diffusion models demonstrate superior training stability and sample quality in comparison to GANs.
Diffusion models \cite{sohl2015deep} are a series of models with a parameterized Markov chain that add noise to gradually degrade images via a forward diffusion process and generate synthetic images from pure noise through a backward denoising process. Denoising diffusion probabilistic models \cite{ho2020denoising} combine diffusion models with denoising score matching and are capable of generating high-fidelity images.
Improved denoising diffusion probabilistic models (IDDPMs) \cite{nichol2021improved} make some modifications based on DDPM, such as learnable variance and cosine variance schedule, to improve log likelihoods while maintaining sample quality.

Introducing conditions to DDPMs would enhance the controllability of image synthesis processes. Those conditions encompass various modalities, including class labels, bounding boxes, semantic masks, and textual contexts, enabling image synthesis tasks such as text-to-image, layout-to-image, semantic-to-image, and image-to-image translation. Conditional DDPMs have advanced in these tasks. ADM \cite{dhariwal2021diffusion} is conditioned on class labels through adaptive group normalization (AdaGN) layer, incorporating conditions into each residual block after group normalization.
Latent diffusion models \cite{rombach2022high} project images from pixel space to latent space and carry out diffusion operations in the latent space, which is computationally efficient. Additionally, LDMs enable multi-modal conditions via cross-attention module. Semantic diffusion models \cite{wang2022semantic} control the denoising process by feeding the semantic mask to the denoising network via spatially adaptive normalization \cite{park2019semantic}. ControlNet \cite{zhang2023adding} learns task-specific conditions to control large pre-trained diffusion models via copying trainable large diffusion model weights.

To incorporate lymph node masks and anatomical structure masks into the conditional diffusion model, we utilize the straightforward yet effective channel-wise concatenation method. Furthermore, we adapt the cross-attention module from LDM \cite{rombach2022high} with specific modifications to further condition the generation of lymph nodes.

\section{Method}

\begin{figure*}[!t]
\centerline{\includegraphics[width=\linewidth]{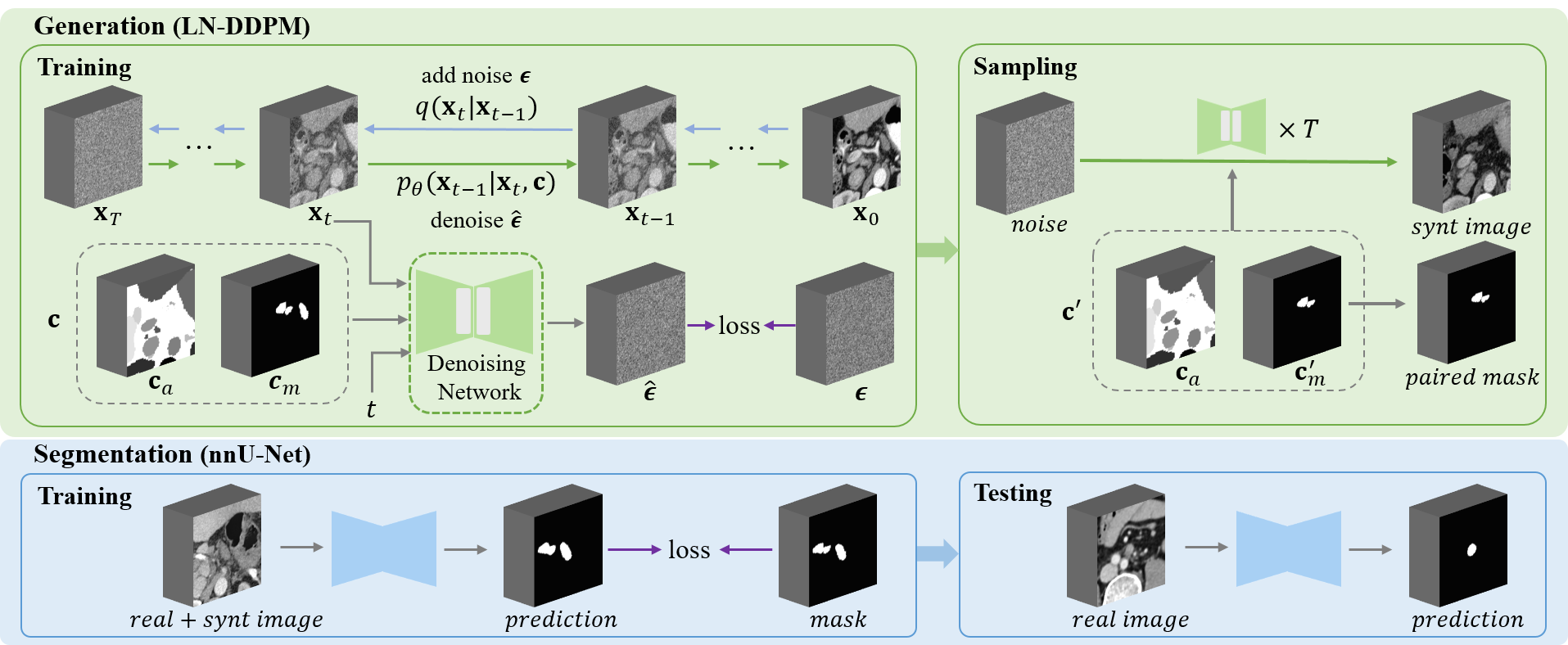}}
\caption{Overview of our proposed pipeline containing LN-DDPM for abdominal lymph node generation and nnU-Net for abdominal lymph node segmentation. The LN-DDPM is conditioned on $\mathbf{c}$, including anatomical structure mask $\mathbf{c}_a$ and lymph node mask $\mathbf{c}_m$, at each denoising step $t$. During training, we compute the loss between the actual added noise $\boldsymbol{\epsilon}$ and the denoising network predicted noise $\boldsymbol{\hat{\epsilon}}$. During sampling, we augment the original condition $\mathbf{c}$ to the transformed condition $\mathbf{c}^{\prime}$, then generate paired abdominal lymph node image and mask for training the downstream segmentation model. The nnU-Net model is trained using both real and synthetic data and tested on real abdominal lymph node images.}
\label{overview}
\end{figure*}

In this section, we provide some preliminaries of diffusion models \cite{sohl2015deep}, then introduce our overall framework containing the LN-DDPM for abdominal lymph node synthesis and the nnU-Net \cite{isensee2021nnu} model for abdominal lymph node segmentation.

\subsection{Preliminaries}
Diffusion models \cite{sohl2015deep} encompass a forward diffusion process that gradually adds noise to images and a reverse denoising process that progressively denoises from random noise to images.
In this paper, leveraging the capabilities of diffusion models, we generate abdominal lymph node images to facilitate the abdominal lymph node segmentation task. These abdominal lymph node images are 3D lymph node patches instead of entire abdominal CT images. This is because the entire abdominal CT images are impractical to generate and contain too much irrelevant background information, whereas 3D lymph node patches contain sufficient information about lymph nodes and their surroundings.

In the diffusion process, given a 3D lymph node patch $\mathbf{x}_0$ that follows real data distribution $q(\mathbf{x})$, for $T$ timesteps, it produces a series of noisy images $\mathbf{x}_1,...,\mathbf{x}_T$. If $T\rightarrow\infty$, $\mathbf{x}_T$ will be pure Gaussian noise. The diffusion process, with variances $\beta_1,...,\beta_T$, can be formulated as:
\begin{equation}
q(\mathbf{x}_t|\mathbf{x}_{t-1})=\mathcal{N}(\mathbf{x}_t;\sqrt{1-\beta_t}\mathbf{x}_{t-1},\beta_t\mathbf{I}).
\label{qt}
\end{equation}

By reparameterization trick, $\mathbf{x}_t$ can be directly derived from $\mathbf{x}_0$ as shown in \eqref{q0}, where $\alpha_t=1-\beta_t$ and $\overline{\alpha}_{t}=\prod_{s=1}^t\alpha_{s}$.
\begin{equation}
q(\mathbf{x}_t|\mathbf{x}_0)=\mathcal{N}(\mathbf{x}_t;\sqrt{\overline{\alpha}_t}\mathbf{x}_0,(1-\overline{\alpha}_t)\mathbf{I})
\label{q0}
\end{equation}

In the reverse process, the diffusion model learns to denoise random Gaussian noise into realistic images through predicting the learnable Gaussian transition with parameters $\theta$:
\begin{equation}
p_\theta(\mathbf{x}_{t-1}|\mathbf{x}_t)=\mathcal{N}(\mathbf{x}_{t-1};\boldsymbol{\mu}_\theta(\mathbf{x}_t,t),\boldsymbol{\Sigma}_\theta(\mathbf{x}_t, t)).
\label{p}
\end{equation}
where the mean $\boldsymbol{\mu}_\theta(\mathbf{x}_t,t)$ is learned by a denoising network $\boldsymbol{\epsilon}_\theta(\mathbf{x}_t,t)$ and the variance $\boldsymbol{\Sigma}_\theta(\mathbf{x}_t, t)=\sigma_t^2\mathbf{I}$, here $\sigma_t$ is often set to a fixed $\beta_t$ or $\tilde{\beta_t}=\frac{1-\overline{\alpha}_{t-1}}{1-\overline{\alpha}_t}\beta_t$. Consequently, $\mathbf{x}_{t-1}$ can be predicted from $\mathbf{x}_t$ via \eqref{xt}, where $\mathbf{z}\sim\mathcal{N}(0,\mathbf{I})$.
\begin{equation}
\mathbf{x}_{t-1}=\frac{1}{\sqrt{\alpha_t}}\left(\mathbf{x}_t-\frac{1-\alpha_t}{\sqrt{1-\overline{\alpha}_t}}\boldsymbol{\epsilon}_\theta(\mathbf{x}_t,t)\right)+\sigma_t\mathbf{z}
\label{xt}
\end{equation}

The objective function for optimizing parameters $\theta$ of the denoising network $\boldsymbol{\epsilon}_\theta(\cdot)$ for noise prediction using $L_1$ loss \cite{dorjsembe2023conditional} is defined as:
\begin{equation}
\mathcal{L}_{DM}=\mathbb{E}_{\mathbf{x}_0,\boldsymbol{\epsilon},t}\left[||\boldsymbol{\epsilon}-\boldsymbol{\epsilon}_\theta(\mathbf{x}_t,t)||\right],
\label{dmloss}
\end{equation}
where $\boldsymbol{\epsilon}\sim\mathcal{N}(0,\mathbf{I})$ and $\mathbf{x}_t=\sqrt{\overline{\alpha}_t}\mathbf{x}_0+\sqrt{(1-\overline{\alpha}_t)}\boldsymbol{\epsilon}$.

\subsection{Overall Framework}
As depicted in Fig. \ref{overview}, the proposed pipeline comprises a generative model for abdominal lymph node generation and a segmentation model for abdominal lymph node segmentation.

The generative model, LN-DDPM, combines global structure conditioning and local detail conditioning with lymph node masks and anatomical structures as conditions. Initially, we train the LN-DDPM using abdominal lymph node images $\mathbf{x}_0$ along with their corresponding lymph node masks $\mathbf{c}_m$ and anatomical structure masks $\mathbf{c}_a$ as conditions $\mathbf{c}$. The denoising network $\boldsymbol{\epsilon}_\theta$ is utilized for noise prediction. We calculate the loss between the predicted noise $\boldsymbol{\hat{\epsilon}}$ and the actual added noise $\boldsymbol{\epsilon}$. Subsequently, we sample synthetic images using the trained LN-DDPM. The transformed conditions $\mathbf{c}^{\prime}$ used for sampling are adapted from real image conditions $\mathbf{c}$ through random transformations. Therefore, we obtain a large quantity of synthetic images with their corresponding masks. 

For the segmentation model, we employ the nnU-Net model for robust abdominal lymph node segmentation. We take advantage of paired real and synthetic abdominal lymph node data for the nnU-Net segmentation model training. The trained nnU-Net segmentation model can further be applied to segment real-world data.

\subsection{Conditioning Signals}

\begin{figure}[!t]
\centerline{\includegraphics[width=0.7\linewidth]{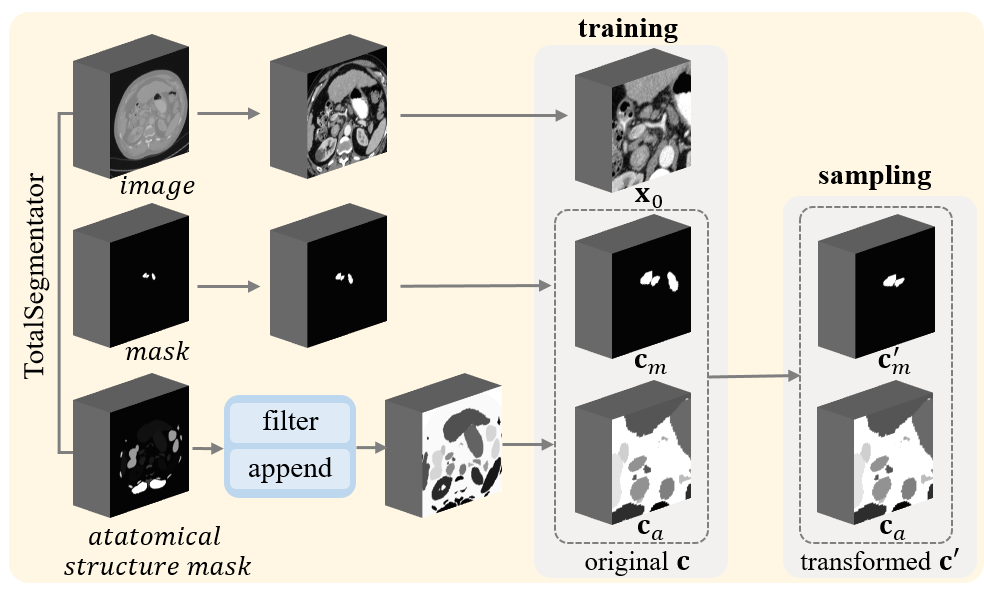}}
\caption{Preparation of conditioning signals. We focus on the lymph node region instead of the entire abdominal cavity. The anatomical structure mask is acquired from TotalSegmentator. The original condition $\mathbf{c}$ is used for diffusion model training and the transformed condition $\mathbf{c}^{\prime}$ is used for diffusion model sampling so as to enhance the conditional diversity of lymph node masks while preserving anatomical structures.}
\label{condition}
\end{figure}

\subsubsection{Lymph node mask.}
In order to assist the downstream task of abdominal lymph node segmentation, it is necessary to generate paired abdominal lymph node images and masks. Hence, abdominal lymph node masks are essential for controlling the diffusion model to generate abdominal lymph node images.

\subsubsection{Anatomical structure mask.}
We generate 3D lymph node patches containing not only lymph nodes but also their surroundings because the abdominal environment where lymph nodes are located is also important. Therefore, except for lymph node masks, we utilize anatomical structure masks to further control the simulation of the realistic abdominal environment. These anatomical structures refer to the anatomical structure masks of abdominal organs and tissues. As shown in Fig. \ref{condition}, given an image, its anatomical structure mask is acquired from TotalSegmentator \cite{wasserthal2023totalsegmentator}, an automated segmentation tool for segmenting anatomical structures in body CT scans. This anatomical structure mask of the entire abdominal cavity is cropped to the lymph node region containing all the lymph nodes with a certain range of expanding in each direction of each axis. The obtained anatomical structure mask is filtered based on whether specific labels belong to abdomen or not. In addition, we further incorporate air label and body label, obtained through thresholding methods, into the anatomical structure mask.

During diffusion model training, we random crop a 3D lymph node patch, i.e., abdominal lymph node image $\mathbf{x}_0$. The corresponding lymph node mask $\mathbf{c}_m$ and anatomical structure mask $\mathbf{c}_a$ are jointly adopted as model condition $\mathbf{c}$. Therefore, the foreground lymph node masks and the background anatomical structure masks reveal the relative position relations between lymph nodes and anatomical structures.

During diffusion model sampling, to obtain diverse sampling conditions, we transform the original condition $\mathbf{c}$ into the transformed condition $\mathbf{c}^{\prime}$ through random transformations. The anatomical structure mask $\mathbf{c}_a$ remains unchanged, whereas the lymph node mask $\mathbf{c}_m$ is transformed into $\mathbf{c}^{\prime}_m$. These random transformations are associated with lymph node shape and lymph node number. We apply random affine or elastic transformations to alter the lymph node shape and random remove a lymph node if multiple lymph nodes are present. We ensure that these lymph nodes do not overlap with other abdominal organs and tissues after random transformations.

\subsection{Conditioning Mechanisms}

\begin{figure}[!t]
\centerline{\includegraphics[width=0.7\linewidth]{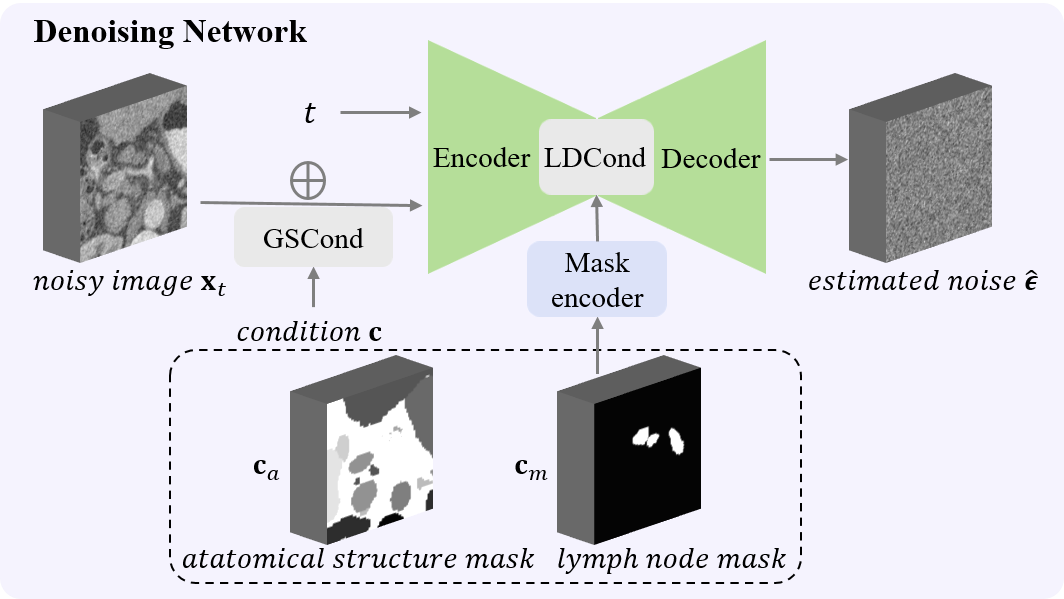}}
\caption{The architecture of the denoising network. The denoising network employs the U-Net architecture. GSCond and LDCond denote global structure conditioning mechanism via channel-wise concatenation and local detail conditioning mechanism via cross-attention, respectively. The mask encoder is used to encode lymph node mask.}
\label{model}
\end{figure}

As illustrated in Fig. \ref{model}, the denoising network $\boldsymbol{\epsilon}_\theta$ of LN-DDPM takes noisy image $\mathbf{x}_t$, current timestep $t$, and condition $\mathbf{c}$ as network inputs and outputs estimated noise $\boldsymbol{\hat{\epsilon}}$. The conditioning signals, anatomical structure mask $\mathbf{c}_a$ and lymph node mask $\mathbf{c}_m$, operate with two kinds of conditioning mechanisms: global structure conditioning and local detail conditioning.

To generate synthetic images that follow actual abdominal anatomical structures, LN-DDPM adopts global structure conditioning mechanism that utilizes both anatomical structure mask and lymph node mask. Global structure conditioning mechanism is implemented by channel-wise concatenation, a straightforward yet effective method. We first convert the multi-label anatomical structure mask $\mathbf{c}_a\in\mathbb{R}^{1\times H\times W\times D}$ into multi-channel anatomical structure mask $\mathbf{c}_a^*\in\mathbb{R}^{C\times H\times W\times D}$. Each channel of the multi-channel anatomical structure mask is the structure of each abdominal organ or tissue. Consequently, the multi-channel anatomical structure mask emphasizes each abdominal organ or tissue separately. Then the lymph node mask $\mathbf{c}_m\in\mathbb{R}^{1\times H\times W\times D}$ and the multi-channel anatomical structure mask $\mathbf{c}_a^*$ are concatenated to form the condition $\mathbf{c}$,
\begin{equation}
\mathbf{c}^{(C+1,H,W,D)}:=\mathbf{c}_a^{*(C,H,W,D)}\oplus\mathbf{c}_m^{(1,H,W,D)}.
\end{equation}
Then the condition $\mathbf{c}$ is concatenated with the noisy image $\mathbf{x}_t\in\mathbb{R}^{1\times H\times W\times D}$ at each denoising timestep $t$ to produce the denoising network input $\tilde{\mathbf{x}}_t$,
\begin{equation}
\tilde{\mathbf{x}}_t^{(C+2,H,W,D)}:=\mathbf{x}_t^{(1,H,W,D)}\oplus\mathbf{c}^{(C+1,H,W,D)}.
\end{equation}
Through global structure conditioning mechanism, LN-DDPM emphasizes the abdominal anatomical structures of lymph nodes and their surroundings.

However, lymph nodes are relatively small compared to larger abdominal organs and tissues, thus the control exerted by global structure conditioning mechanism over lymph nodes is relatively weak. To synthesize more realistic abdominal lymph nodes and better support the downstream abdominal lymph node segmentation task, we expect the generation model to better capture lymph node characteristics. Therefore, LN-DDPM incorporates local detail conditioning mechanism to enhance control over lymph nodes. The local detail conditioning mechanism is implemented using the spatial transformer \cite{vaswani2017attention} as exploited in LDM \cite{rombach2022high}, comprising a self-attention module, a cross-attention module, and a feed-forward neural network. Through cross-attention module, conditions are injected. Initially, we pass lymph node mask $\mathbf{c}_m$ through a mask encoder $\boldsymbol{\tau}_\theta$, which comprises several convolution operations, to obtain the mask representation. Subsequently, the mask representation is carried out cross-attention with the intermediate representations of the U-Net \cite{ronneberger2015u}. To highlight lymph node characteristics, the local detail conditioning mechanism operates exclusively on the lymph nodes themselves, forwarding the spatial transformer output of lymph nodes via a residual connection \cite{he2016deep} rather than the holistic output. 

Our conditional diffusion model, LN-DDPM, utilizes anatomical structure mask $\mathbf{c}_a$ and lymph node mask $\mathbf{c}_m$ as condition $\mathbf{c}$, with the denoising network represented as $\boldsymbol{\epsilon}_\theta(\mathbf{x}_t,\mathbf{c},\boldsymbol{\tau}_\theta(\mathbf{c}_m),t)$. The objective function using $L_1$ loss \cite{dorjsembe2023conditional} is formulated as:
\begin{equation}
\mathcal{L}_{CDM}=\mathbb{E}_{\mathbf{x}_0,\mathbf{c},\boldsymbol{\epsilon},t}\left[||\boldsymbol{\epsilon}-\boldsymbol{\epsilon}_\theta(\mathbf{x}_t,\mathbf{c},\boldsymbol{\tau}_\theta(\mathbf{c}_m),t)||\right].
\label{cdmloss}
\end{equation}

\section{Experiments}
In this section, we describe the experimental settings, including medical image datasets, implementation details, baseline methods, and evaluation metrics.

\subsection{Datasets}
In our study, we leverage both publicly available and proprietary datasets to evaluate the performance of our proposed method. Specifically, we utilize the Abdominal Lymph Node (ABD-LN) dataset and the Colorectal Cancer Lymph Node Metastasis (CRC-LNM) dataset to conduct comprehensive evaluations on abdominal lymph node generation and segmentation tasks.

\subsubsection{ABD-LN dataset.}
Sourced from The Cancer Imaging Archive (TCIA) \cite{clark2013cancer} and described by Roth \emph{et al.} \cite{roth2014new}, the ABD-LN dataset comprises 86 CT images from 86 patients, totaling 595 identified abdominal lymph nodes. This dataset provides a broad spectrum of instances for analyzing the characteristics and variations of abdominal lymph nodes.

\subsubsection{CRC-LNM dataset.}
The CRC-LNM dataset, obtained from the First Hospital of China Medical University, includes 52 patients who underwent surgery for colorectal cancer with oligometastatic lymph nodes. The determination of metastatic lymph node was made by experienced pathologists and radiologists, through a meticulous spatial correspondence between postoperative station-specific lymph node pathology reports and preoperative CT images. The dataset includes 52 CT images of 145 lymph nodes with confirmed metastasis through both pathology and imaging, ensuring high-quality imaging and reliability of results.

We retain 17 cases from the ABD-LN dataset and 10 cases from the CRC-LNM dataset for the segmentation model testing. The remaining 69 cases from the ABD-LN dataset and 42 cases from the CRC-LNM dataset are jointly utilized for the generative model and the segmentation model training, with 88 cases used for training and 23 cases used for validation.

\subsection{Implementation Details}
We implement our method in Python 3.8 and PyTorch 1.12.0. All the experiments are carried out on NVIDIA GeForce RTX 3090 GPU with 24\,GB memory.

\subsubsection{Generative model.}
We apply our proposed LN-DDPM as the generative model. The code structure is based on the Medical Diffusion \cite{khader2023denoising}. And the denoising network architecture is based on the 3D U-Net model from LDM \cite{rombach2022high}, with some modifications to implement our proposed method. Spatial transformer is performed at spatial resolution of 16 × 16 × 16. For data processing, we first crop the entire lymph node region of each abdominal CT with 10\,cm expansions in each direction of each axis, then resample the voxel spacing to 1.0\,mm × 1.0\,mm × 1.0\,mm. Afterward, window truncation is performed by clipping each abdominal CT into range [-120, 240]. Finally, these abdominal CT images are normalized to [-1, 1]. The multi-label anatomical structure mask has 14 labels and the multi-channel anatomical structure mask has 14 channels.
In addition, the patch size used for training and sampling is 128 × 128 × 128. The diffusion timesteps $T$ is set to 300 with cosine noise schedule. We utilize $L_1$ loss and Adam optimizer for training with $\beta_1$ of 0.9, $\beta_2$ of 0.999, a learning rate of $10^{-4}$, and a batch size of 1. The exponential moving average decay is set to 0.995. We train our network for 45k iterations.

\subsubsection{Segmentation model.}
We employ 3D nnU-Net \cite{isensee2021nnu} as the segmentation model. In this paper, we perform abdominal lymph node segmentation in the local region, which means that given an image, we first localize the entire lymph node region and then crop this region with surroundings expanded. We expand 10\,cm outside the lymph node region in each direction of each axis during training and 5\,cm during testing.
We utilize two kinds of training strategies during training phase. The first training strategy exclusively uses synthetic data to train the nnU-Net model. The second training strategy utilizes both synthetic data and real data to train the nnU-Net model. After training, we test segmentation models on unseen real data.

\begin{figure*}[!t]
\centerline{\includegraphics[width=\linewidth]{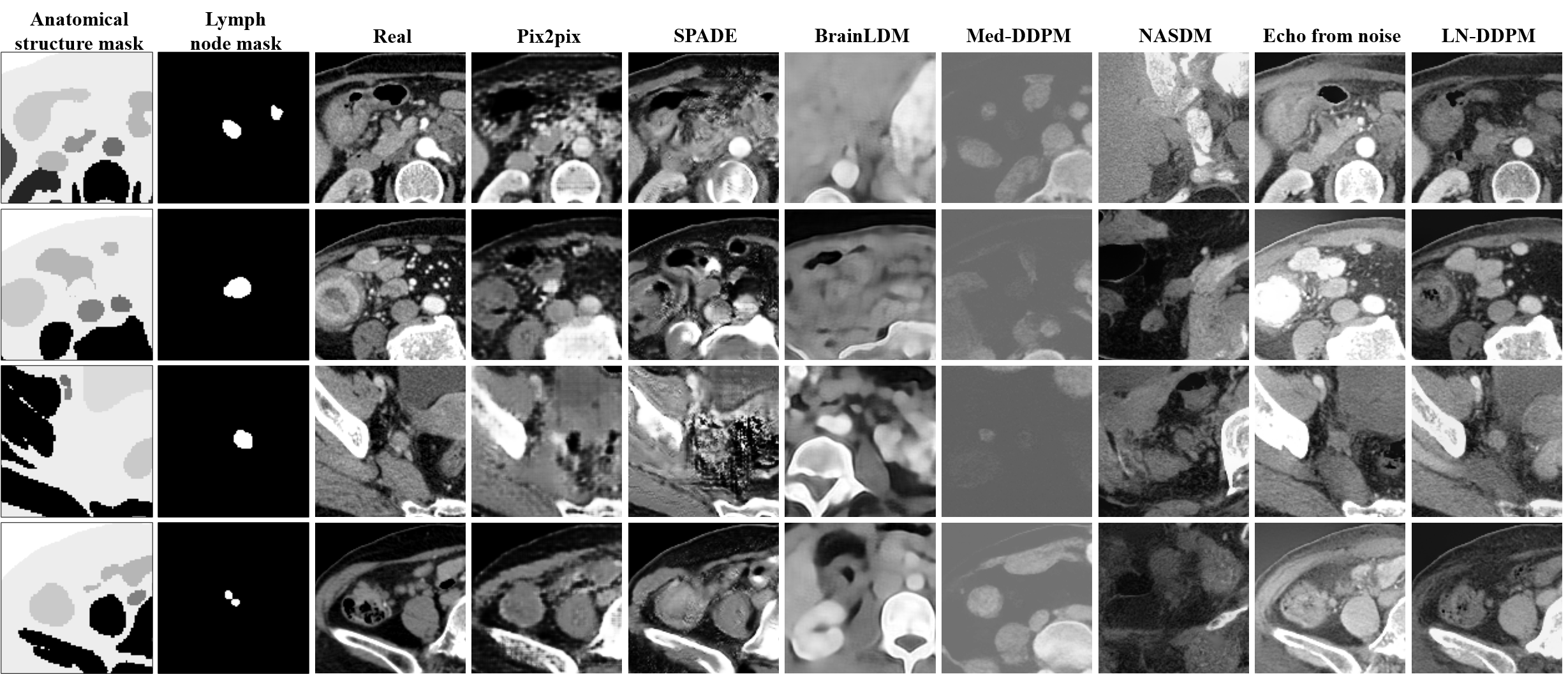}}
\caption{Visualization of generation results. The first column and the second column list sampling conditions for LN-DDPM to synthesize abdominal lymph node images. The third column lists real abdominal lymph node images. Methods conditioned on anatomical structures generate more realistic abdominal lymph node images. Besides, LN-DDPM generates lymph node images with more clear boundaries.}
\label{gen}
\end{figure*}

\subsection{Baseline Methods}
To demonstrate the effectiveness of our proposed method, we compare it against two GAN-based methods and four diffusion-based methods for evaluating the performance of abdominal lymph node generation and segmentation.

\subsubsection{GAN-based methods.}
Pix2pix \cite{isola2017image} is an image-to-image translation method, and we use semantic masks as the source domain and abdominal images as the target domain. SPADE \cite{park2019semantic} is a semantic image synthesis method, making use of semantic masks through spatially-adaptive normalization. For these two GAN-based methods, we adapt their official 2D architecture to 3D in order to suit our task. Both methods use anatomical structure mask and lymph node mask as conditions.

\subsubsection{Diffusion-based methods.}
Pinaya \emph{et al.} \cite{pinaya2022brain} use covariable such as age, sex, and brain structure volumes as conditions to generate 3D brain MRIs via LDM. We refer to this method as BrainLDM \cite{pinaya2022brain} for brevity. We use the implementation provided by MONAI Generative Models \cite{pinaya2023generative}. To assist the abdominal lymph node segmentation task, we consider the lymph node mask as model condition and replace the frozen CLIPTextModel \cite{radford2021learning} with a trainable condition encoder.
Med-DDPM \cite{dorjsembe2023conditional} synthesizes brain tumor MRIs using brain and tumor masks. Thus, we control the diffusion model using lymph node mask and coarse anatomical structure mask, specifically the body mask obtained by thresholding operations.
NASDM \cite{shrivastava2023nasdm} is a nuclei pathology synthesis model via SDM. We change its original 2D structure to 3D and replace the nuclei mask and edge with the lymph node mask and edge as network conditions.
Echo from noise \cite{stojanovski2023echo} also adopts SDM to synthesize cardiac ultrasounds using semantic masks. We also change its 2D structure to 3D and use anatomical structure mask and lymph node mask to control the diffusion model.

\subsection{Evaluation Metrics}
We utilize metrics widely used to evaluate segmentation performance, including dice similarity coefficient (DSC), intersection over union (IOU), Recall, Precision, and average symmetric surface distance (ASSD), as demonstrated in previous studies \cite{li2020deep,wang2017central}.
In addition, to better evaluate the segmentation results from the perspective of lymph node detection, NodeRecall is employed for further assessment. We define that given predictions and ground-truths, a lymph node is considered detected if DSC between any prediction and its ground-truth is greater than 0.1.

\section{Results}
\subsection{Generation Results}
We evaluate the 3D abdominal lymph node CT generation performance of baseline methods against our proposed method.

\subsubsection{From the perspective of generation.}
Fig. \ref{gen} visualizes the generation results of baseline methods and the proposed method under their respective sampling conditions. The third column displays the real abdominal lymph node images. Without anatomical structure mask as model condition, BrainLDM, Med-DDPM, and NASDM struggle to capture abdominal anatomical structure distributions. The abdominal lymph nodes generated by these methods are difficult to distinguish from other abdominal regions. Pix2pix, SPADE, Echo from noise, and LN-DDPM generate abdominal lymph node images that follow the realistic abdominal anatomical structures. The abdominal lymph node images synthesized by Pix2pix and SPADE, appear fuzzy and artificial compared to Echo from noise and LN-DDPM. Moreover, LN-DDPM generates abdominal lymph node images with more clear boundaries, especially the lymph nodes.

\subsubsection{From the perspective of segmentation.}
To further represent the quality of abdominal lymph node generation, we refer to the second part of results in Table \ref{comseg}. We only use synthetic data to train the nnU-Net model, thus enabling the evaluation of generation performance directly and distinctly. Our proposed LN-DDPM not only outperforms other baseline methods dramatically but also achieves superior DSC and comparable IOU in comparison to training with real abdominal lymph node images. Thus, LN-DDPM demonstrates its capability of generating realistic abdominal lymph node images.

\begin{table}[!t]
\centering
\caption{Quantitative results on abdominal lymph node segmentation under different training strategies. We adopt different generative methods for synthetic image generation and the nnU-Net model for segmentation.}
\resizebox{\linewidth}{!}{
\begin{tabular}{llcccccc}
\toprule
Strategy & Method & DSC ↑ & IOU ↑ & Recall ↑ & Precision ↑ & ASSD ↓ & NodeRecall ↑ \\
\hline
Real & - & 0.5273 & 0.3960 & 0.4936 & 0.6447 & 9.0027 & 0.6405 \\
\hline
Synt
& Pix2pix & 0.0898 & 0.0582 & 0.1097 & 0.0952 & 23.8107 & 0.1710 \\
& SPADE & 0.3444 & 0.2452 & 0.3475 & 0.4081 & 16.6340 & 0.5525 \\
& BrainLDM & 0.0700 & 0.0398 & 0.2482 & 0.0466 & 15.7442 & 0.3029 \\
& Med-DDPM & 0.0850 & 0.0573 & 0.0811 & 0.1292 & 25.7986 & 0.1837 \\
& NASDM & 0.4728 & 0.3478 & 0.5349 & 0.5544 & 9.4852 & 0.6646 \\
& Echo from noise & 0.5217 & 0.3815 & 0.5771 & 0.5307 & 9.7072 & 0.7329 \\
& LN-DDPM & \textbf{0.5376} & \textbf{0.3958} & \textbf{0.5830} & \textbf{0.5547} & \textbf{8.8986} & \textbf{0.7486} \\
\hline
Real + Synt
& Pix2pix & 0.4838 & 0.3439 & 0.4848 & 0.5919 & 8.0297 & 0.6170 \\
& SPADE & 0.5012 & 0.3637 & 0.5147 & 0.5591 & 8.7973 & 0.6425 \\
& BrainLDM & 0.1495 & 0.0975 & 0.2729 & 0.1696 & 13.8108 & 0.3748 \\
& Med-DDPM & 0.5186 & 0.3922 & 0.5075 & 0.5942 & 7.8913 & 0.6424 \\
& NASDM & 0.5323 & 0.3955 & 0.5336 & 0.6396 & 7.5384 & 0.6824 \\
& Echo from noise & 0.5460 & 0.4056 & 0.5432 & 0.6384 & 8.2362 & 0.6855 \\
& LN-DDPM & \textbf{0.5655} & \textbf{0.4200} & \textbf{0.5591} & \textbf{0.6451} & \textbf{7.1851} & \textbf{0.7054} \\
\toprule
\end{tabular}
}
\label{comseg}
\end{table}

\begin{figure}[!t]
\centerline{\includegraphics[width=0.7\linewidth]{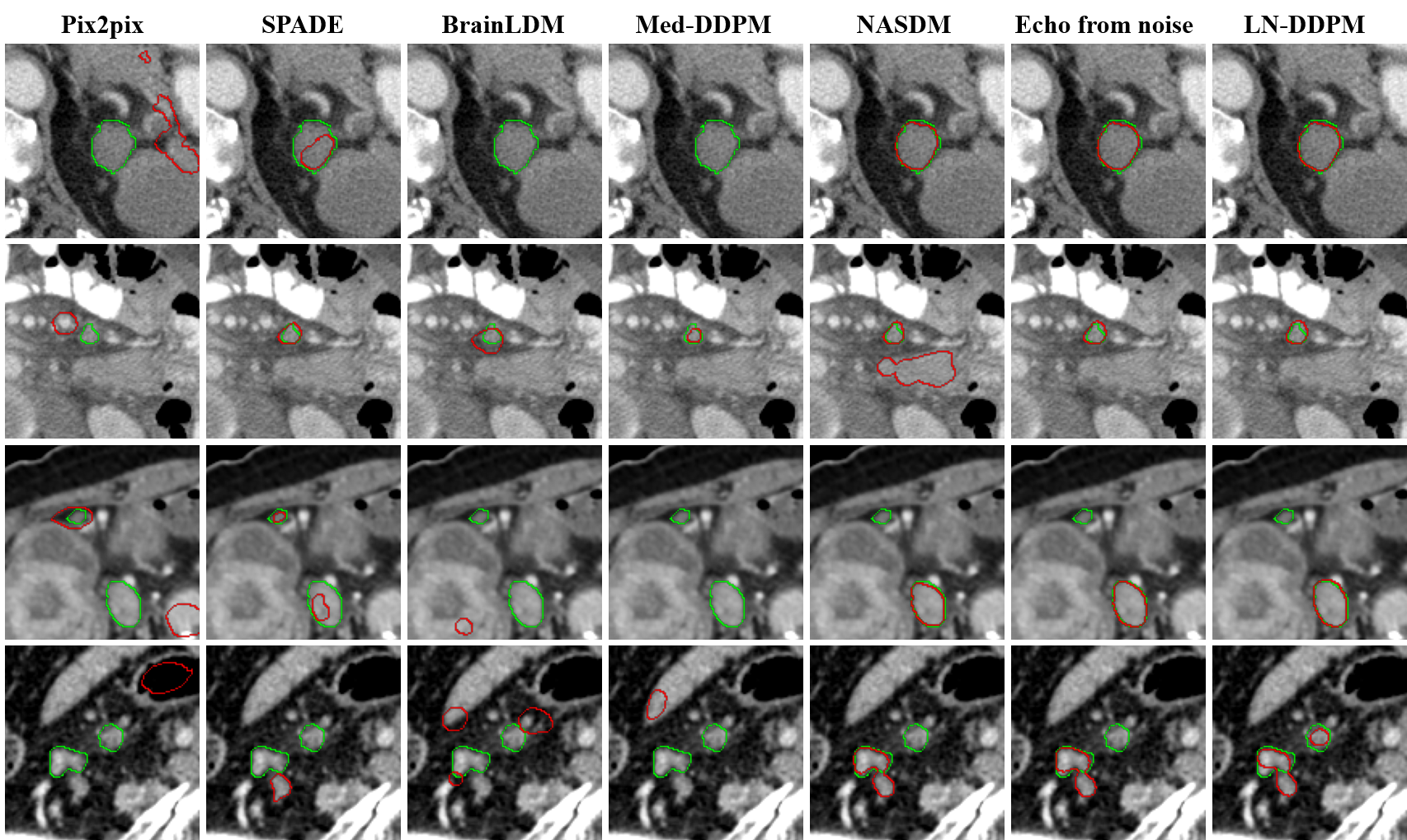}}
\caption{Visualization of segmentation results when training exclusively with synthetic data. The green and red colors represent ground-truths and predictions of abdominal lymph nodes, respectively. LN-DDPM exhibits superior segmentation performance.}
\label{seg}
\end{figure}

\subsection{Segmentation Results}
We conduct experiments to evaluate the effectiveness of baseline methods and our proposed method in facilitating the downstream abdominal lymph node segmentation task.

Table \ref{comseg} displays the abdominal lymph node segmentation results of different generative methods under different training strategies. The number of synthetic abdominal lymph node images generated using different generative methods is ten times that of those 88 real cases. The first row displays the segmentation results using 88 real cases from ABD-LN dataset and CRC-LNM dataset to train the nnU-Net model, where DSC, IOU, Recall, Precision, ASSD, and NodeRecall are 0.5273, 0.3960, 0.4936, 0.6447, 9.0027\,mm, and 0.6405, respectively.

\subsubsection{Training only with synthetic data.}
When training the nnU-Net model exclusively with synthetic data, our proposed LN-DDPM gains better DSC, Recall, ASSD, and NodeRecall of 0.5376, 0.5830, 8.8986\,mm and 0.7486, respectively, while maintaining a similar IOU of 0.3958 compared to training with real abdominal lymph node images. However, the Precision is inferior to training with real abdominal lymph node images primarily because those synthetic images are lymph node region images and lack richer information about abdominal environment.
The visualized segmentation results of different generative approaches when training with only synthetic data are depicted in Fig. \ref{seg}. LN-DDPM is capable of discovering the majority of abdominal lymph nodes with fewer false positives and achieving better segmentation accuracy. Notably, in the last row, only LN-DDPM segments the upper right lymph node. However, LN-DDPM fails to detect the small upper left lymph node in the third row.

\subsubsection{Training with both real and synthetic data.}
We take advantage of both real and synthetic data to train the nnU-Net model. Through cooperation with real images, the segmentation metrics exhibit considerable improvements compared to training exclusively with synthetic images. LN-DDPM obtains the state-of-the-art results with DSC of 0.5655, IOU of 0.4200, Recall of 0.5591, Precision of 0.6451, ASSD of 7.1851\,mm, and NodeRecall of 0.7054.
In summary, these results highlight the effectiveness of utilizing synthetic data generated by LN-DDPM for training abdominal lymph node segmentation model. 

\begin{table}[!t]
\centering
\caption{Ablation study on model conditioning signals and conditioning mechanisms. AS denotes anatomical structures.}
\resizebox{\linewidth}{!}{
\begin{tabular}{ccccccccc}
\toprule
\makecell{GSCond \\w/ LN mask} & \makecell{GSCond \\w/ AS} & LDCond & DSC ↑ & IOU ↑ & Recall ↑ & Precision ↑ & ASSD ↓ & NodeRecall ↑ \\
\hline
\Checkmark & & & 0.3421 & 0.2407 & 0.4281 & 0.3215 & 13.4891 & 0.5946 \\
\Checkmark & \Checkmark & & 0.5255 & 0.3898 & 0.5680 & 0.5539 & 9.4574 & 0.7209 \\
\Checkmark & \Checkmark & \Checkmark & \textbf{0.5376} & \textbf{0.3958} & \textbf{0.5830} & \textbf{0.5547} & \textbf{8.8986} & \textbf{0.7486} \\
\toprule
\end{tabular}
}
\label{condseg}
\end{table}

\subsection{Ablation Studies}
\subsubsection{Conditioning signals and mechanisms.}
To assess the effectiveness of each component of LN-DDPM, we conduct ablation studies on its conditioning signals and mechanisms. We separate LN-DDPM into three components and integrate them progressively. These three components include global structure conditioning with lymph node mask, global structure conditioning with anatomical structure mask, and local detail conditioning. The segmentation results of LN-DDPM under different conditioning signals and mechanisms are presented in Table \ref{condseg}. Combining these three components leads to a significant enhancement in segmentation performance. Through the combination of the local detail conditioning mechanism, DSC and IOU increase by approximately one point, Recall and NodeRecall increase by around two points, while ASSD decreases.

\subsubsection{Number of synthetic images.}
We evaluate the impact of using different number of synthetic images to assist the downstream abdominal lymph node segmentation task as shown in Table \ref{syntnum}. We generate synthetic abdominal lymph node images at 5×, 10×, and 20× the size of real abdominal lymph node images and train the nnU-Net model using both real and synthetic data. With 10× synthetic data, DSC, IOU, and ASSD are better. With 5× synthetic data, Recall and NodeRecall are higher, while Precision remains nearly constant. However, segmentation performance decreases when using 20× synthetic data. This may because when training with both real and synthetic data, too many synthetic images may lead to the neglect of real images.

\begin{table}[!t]
\caption{Ablation studies.}
\centering
\subtable[Ablation study on the number of synthetic images.]
{
\begin{tabular}{ccccccc}
\toprule
\#Synt & DSC & IOU & Recall & Precision & ASSD & NodeRecall \\
\hline
5× & 0.5583 & 0.4150 & \textbf{0.5655} & \textbf{0.6456} & 7.6136 & \textbf{0.7167} \\
10× & \textbf{0.5655} & \textbf{0.4200} & 0.5591 & 0.6451 & \textbf{7.1851} & 0.7054 \\
20× & 0.5345 & 0.3943 & 0.5580 & 0.5870 & 8.3481 & 0.7118 \\
\toprule
\end{tabular}
\label{syntnum}
}
\subtable[Ablation study on the sampling conditions.]
{
\begin{tabular}{cccccccc}
\toprule
Trans. & DSC & IOU & Recall & Precision & ASSD & NodeRecall \\
\hline
\XSolidBrush & 0.5227 & 0.3807 & 0.5722 & 0.5379 & 9.8697 & 0.7468 \\
\Checkmark & \textbf{0.5376} & \textbf{0.3958} & \textbf{0.5830} & \textbf{0.5547} & \textbf{8.8986} & \textbf{0.7486} \\
\toprule
\end{tabular}
\label{convert}
}
\subtable[Ablation study on the attention modes in the spatial transformer.]
{
\begin{tabular}{cccccccc}
\toprule
Mode & DSC & IOU & Recall & Precision & ASSD & NodeRecall \\
\hline
LD & \textbf{0.5376} & \textbf{0.3958} & \textbf{0.5830} & 0.5547 & \textbf{8.8986} & 0.7486 \\
LA & 0.5280 & 0.3925 & 0.5753 & \textbf{0.5573} & 9.0464 & 0.7492 \\
GA & 0.5156 & 0.3765 & 0.5760 & 0.5141 & 9.9408 & \textbf{0.7568} \\
\toprule
\end{tabular}
\label{attnway}
}
\end{table}

\subsubsection{Sampling conditions.}
To illustrate the impact of condition transformations during synthesizing abdominal lymph node images, we train the nnU-Net model using synthetic images sampled from both original and transformed conditions. The segmentation results shown in Table \ref{convert} demonstrates obvious improvements, indicating that condition transformations during sampling are effective in synthesizing diverse abdominal lymph node images for training the downstream segmentation model.

\subsubsection{Attention modes.}
To focus more on lymph nodes, the local detail conditioning mechanism operates solely on the lymph nodes themselves. To elaborate the effectiveness of local detail conditioning mechanism, we compare with local attention, which modifies attention operations within the spatial transformer to be local via a lymph node mask, and global attention, which maintains original attention operations. Table \ref{attnway} indicates the results, where LD, LA, and GA refer to local detail conditioning, local attention, and global attention, respectively. Although NodeRecall of local detail conditioning mechanism is slightly lower than that of global attention, it attains significantly better results for other metrics. Besides, local detail conditioning mechanism demonstrates superior DSC, IOU, Recall, and ASSD than local attention, while obtaining comparable Precision and NodeRecall.

\subsection{Limitations}
Our proposed LN-DDPM utilizes anatomical structure masks as model conditions. The anatomical structure masks rely on the segmentation results from TotalSegmentator \cite{wasserthal2023totalsegmentator}, rather than elaborate labeling of organs and tissues. Therefore, the anatomical structure masks are not entirely accurate and serve as rough guidance to abdominal anatomical structures. In addition, due to the limited computational resources, LN-DDPM generates 3D CT images of lymph node regions rather than the entire abdominal cavity. Additionally, the sampling speed of diffusion models is slower than other generative models. Therefore, future research endeavors can investigate methods such as latent diffusion models to generate larger images while accelerating sampling speed, but it is crucial to balance sampling speed and sample quality.

\section{Conclusion}
This paper presents a pipeline that integrates the LN-DDPM for lymph node generation and the nnU-Net model for lymph node segmentation. Using LN-DDPM, we synthesize a variety of abdominal lymph node images to enhance the downstream segmentation task for more accurate lymph node segmentation. Our proposed LN-DDPM utilizes lymph node masks and anatomical structures as model conditions and employs global structure conditioning and local detail conditioning as conditioning mechanisms. Moreover, through extensive experiments on the abdominal lymph node datasets, we demonstrate that LN-DDPM surpasses other generative methods in synthesizing abdominal lymph node images and enhancing the abdominal lymph node segmentation task.

\section*{Acknowledgments}
This work was supported in part by the National Natural Science Foundation of China (82203199) (Hanyu Chen).

\bibliographystyle{splncs04}
\bibliography{ref}

\end{document}